\documentclass[pra,preprint,showpacs]{revtex4}

\usepackage{graphicx}

\newcommand{\rf}[1]{(\ref{#1})}

\newcommand{\beq}{\begin{equation}}
\newcommand{\eeq}{\end{equation}}
\newcommand{\beqr}{\begin{eqnarray}}
\newcommand{\eeqr}{\end{eqnarray}}

\newcommand{\lb}[1]{\label{#1}}

\newcommand{\bc}{\begin{center}}
\newcommand{\ec}{\end{center}}
\newcommand{\ct}[1]{\cite{#1}}
\newcommand{\bi}[1]{\bibitem{#1}}

\begin{document} 

\title{Conditional quantum logic  using two atomic qubits}

\author{I.E.~Protsenko$^{1,2,3}$, G. Reymond$^{1}$, N. Schlosser$^{1}$ and P.~Grangier$^{1}$} 
\affiliation{$^1$ Laboratoire Charles Fabry de l'Institut  d'Optique, UMR 8501 du CNRS, F91403 Orsay, France \\
$^2$ Lebedev Physics Institute, Leninsky Prospect 53, Moscow, Russia\\
$^3$ Scientific Center of Applied Research, JINR, Dubna, Russia}

\vskip 0.5 cm

\begin{abstract}

\noindent In this paper we propose and analyze a feasible 
scheme where the detection of
a single scattered photon from two trapped atoms or ions
performs a conditional unitary operation on two qubits. As examples we consider
the preparation of all four Bell states, the reverse operation that is
a Bell measurement, and a CNOT gate. We study the effect of atomic motion and 
multiple scattering, by evaluating Bell inequalities violations, 
and by calculating the CNOT gate fidelity.

\vskip 0.5 cm


\end{abstract}

\maketitle

\section{Introduction}
	
Implementing a quantum controlled-not (CNOT) gate is a key step in present attempts towards quantum computation
\ct{1a,1b}.  Many different schemes for CNOT gates have been proposed \ct{2a,2b,2c,2d,2e}, and most of them require a
strong quantum interaction between the particles that are used to carry the logical qubits.  In practice, the quantum
interaction is very often perturbed by classical noises, such as the individual motion of neutral atoms in a
laser-induced potential well \ct{3}, or the collective motion of ions in a Paul trap \ct{4}. Though order-of-magnitude
estimations show that most of the CNOT gate schemes may be realized in principle, detailed analysis discovers many
difficulties in eliminating all sources of classical noise for given experimental conditions.  For instance, the
perturbations due to thermal photons, photo-ionization, spontaneous emission..., make that the conditions for a fast
CNOT gate operation through transient excitation to Rydberg states are only marginally satisfied \ct{3}.  If one looks
at cavity-induced atom-atom coupling (``cavity-assisted collisions" \ct{coll})
in the optical domain, our estimations show that most schemes for cavity-enhanced
coupling between the particles reliably works when $g^2/(\kappa\gamma) > 10^3$, where $g$ is a cavity mode-atom coupling
constant, and $\kappa$ and $\gamma$ are respectively the cavity and the spontaneous emission damping rates. Though it is
possible in principle to reach high value of $g^2/(\kappa\gamma)$ \ct{5}, putting together very small high finesse
cavities and reliable traps is far from straightforward.  These considerations encourages us to look for
``non-traditional" CNOT gate schemes, which do not require a direct interaction between particles, but rather use an
interference effect and a measurement-induced state projection to create the desired operation \ct{klm}.  
It was proposed in \ct{5a,6} to create an entangled state of two atoms simply from the detection of a photon,
spontaneously emitted by one of the atoms in such a way that the emitting atom can't be recognized.  In such a scheme there
is no direct interaction between atoms, and in principle the atoms can even be located very far from each other.

In this paper we propose to extend the ideas of \ct{5a}, \ct{6}, to realize a full quantum CNOT gate, or a Bell-state
measurement, or more generally to implement {\it conditional unitary operations}. Our scheme will be based on an
experimental setup using two atoms in two neighboring microscopic dipole traps \ct{3,7}, but it can be readily applied
to other systems. In Section 1 we will describe how to realize a conditional unitary transformation that maps the four
factorized states of two qubits onto the four maximally entangled Bell's states. Since a convenient experimental
signature of entanglement is the violation of Bell's inequalities (BI) \ct{8}, we will evaluate the result of a test of
BI on the ``transformed" pair of qubits, taking into account imperfections due to the motion of atoms (Section 2) and to
the spontaneous emission of two photons by two atoms (Section 3).  BI measurements are studied quantitatively in Section
4. In Section 5 we describe a CNOT gate based on the Bell's states created by the procedure of Section 1, and we
calculate the fidelity of this gate, taking into account the motion of the atoms in the traps and the possible
spontaneous emission of two photons. Finally we discuss these results and suggest developments of the proposed scheme.

\section{Preparing four orthogonal Bell's states}

We consider two atoms $i=1,2$, trapped in two separate dipole traps, and prepared in one of two states $\left|e\right>_i$ or
$\left|g\right>_i$ of the ground state hyperfine structure. We represent four initial states
of the two-atom system as a vector-column
\beq
	\{\left|\alpha\beta\right>\} = \{\left|gg\right>,\left|ge\right>,\left|eg\right>, \left|ee\right>\},
	\hspace{1cm} \alpha,\beta = e,g. \lb{p1}
\eeq
Each atom can be excited to one of upper states $\left|e'\right>_i$ or $\left|g'\right>_i$ by resonant $\sigma$
polarized laser fields of Rabi frequencies $\Omega_{gi}$, $\Omega_{ei}$, as shown in Fig.1. The fields are weak, so that the
probability to excite both atoms is much smaller than the probability to excite only one atom.  An excited atom may emit
spontaneously a photon, with the wave vector ${\bf k}$ and certain polarization, on $\pi$-polarized $\left|g'\right>_i
\rightarrow \left|g\right>_i$ or $\left|e'\right>_i \rightarrow \left|e\right>_i$ transitions. Occasionally a photon
passes through the optical system shown in Fig.2 and it is registered by the photo-detector.  We assume that the
polarizer $P$ transmits only $\pi$-polarized photons, and thus $\sigma$ polarized photons emitted on the
$\left|e'\right>_i \rightarrow \left|g\right>_i$ and $\left|g'\right>_i \rightarrow \left|e\right>_i$ transitions will
not be registered.

%
\begin{figure}[h]
\center{
\includegraphics[width=8cm]{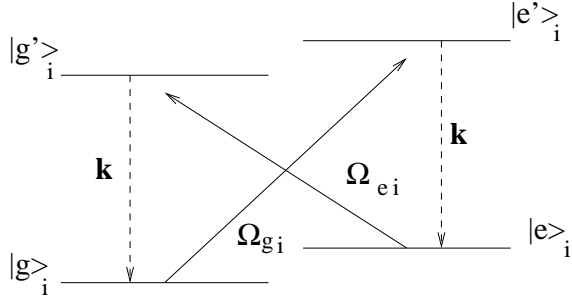} }\\
\caption{ \small Scheme of the relevant atom transitions}
\end{figure}
%
%
\begin{figure}[h]
\center{
\includegraphics[width=8cm]{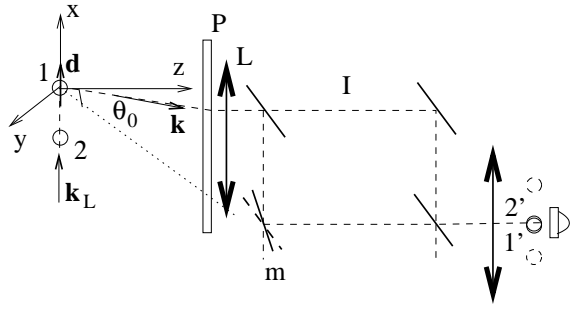} }\\
\caption{ \small Proposed scheme for conditional quantum logic. Atoms 1, 2 are placed in a focal plane of
the input lens $L$ of the optical system. The atom pair is excited by the laser field with the wave vector ${\bf k}_L$,
circularly polarized in the yz-plane, and emits a photon with the wave vector ${\bf k}$ on the x-polarized transition. 
The polarizer $P$ selects x-polarized photons, that are transmitted through an interferometer $I$ towards a
photo-detector. A mirror
$m$ of the interferometer is tilted, so that the images 1', 2' of the two atoms coincide on the photo-detector. 
The unit vector in the direction of a $\pi$-polarized atomic dipole is denoted as ${\bf d}$, and
$\theta_0$ is the aperture angle of the lens $L$.}
\label{setup}
\end{figure}
%
%

After the excitation, the wave function of two atoms is changed from $\left|\alpha\beta\right>$ to
$\left|\Psi_{\alpha}\right>_1\left|\Psi_{\beta}\right>_2$
\beq
	\left|\Psi_e\right>_i = \left|e\right>_i + b\left|g'\right>_i
	e^{i({\bf k}_e\delta{\bf r}_i + \varphi_{ei}) }, \hspace{0.5cm}
	\left|\Psi_g\right>_i = \left|g\right>_i + b\left|e'\right>_i
	e^{({\bf k}_g\delta{\bf r}_i + \varphi_{gi})},  \hspace{0.5cm}
	 \varphi_{\alpha i} = {\bf k}_{\alpha}{\bf r}_i + \varphi_{\alpha i}^0 ,  \lb{1}
\eeq
where $\delta {\bf r}_i$ describes fluctuations in the position of atom $i$ near the equilibrium due to the motion of
the atom in the trap, ${\bf r}_i$ is the atom position at equilibrium, ${\bf k}_{e,g}$ are the wave
vectors of the laser field resonant to either $e\rightarrow g'$ or $g\rightarrow e'$ transitions,
$\varphi_{\alpha i}^0$ is the phase of the laser field $\Omega_{\alpha i}$, $b \ll 1$ is a real constant.

The registration of a photon means that the wave function $\left|\Psi_{\alpha}\right>\left|\Psi_{\beta}\right>$ is
projected to a Bell state $\left|B_{\alpha\beta}\right> = \left|{\bf k}\right>\hat{B}\left|\alpha\beta\right>$ where
$\left|{\bf k}\right>$ is the state of the field with one spontaneously emitted photon. For example,
$\left|\Psi_g\right>\left|\Psi_g\right>$ is projected to
\beq
	\left|{\bf k}\right>\hat{B}\left| gg \right> = \left|{\bf k}\right>\left|B_{gg}\right>,
	\hspace{0.5cm} \left|B_{gg}\right>  = (1/\sqrt{2})
	\left\{\left|ge\right>e^{i[{\bf q}_g\delta{\bf r}_2 + kl_2({\bf k}) + \varphi_{g2}]} +  
	\left|eg\right>e^{i[{\bf q}_g\delta{\bf r}_1 + kl_1({\bf k}) + \varphi_{g1}]}\right\},	\lb{p0}
\eeq
where ${\bf q}_{\alpha} = {\bf k}_{\alpha} - {\bf k}$, $k = |{\bf k}|$ and $l_i({\bf k})$ is the optical length which a
photon travels through the optical system towards the photo-detector.  The optical system is set in such a way, that
images 1', 2' of atoms 1 and 2 perfectly coincide on the photo-detector. This means that $l_i({\bf k})$ is the same for
all registered photons, and therefore $kl_i({\bf k}) = kl_i$. Introducing the vector-column $\{
\left|B_{\alpha\beta}\right>\} = \{ \left|B_{gg}\right>, \left|B_{ge}\right>, \left|B_{eg}\right>,
\left|B_{ee}\right>\}$ of the Bell states we can express them in terms of the initial states \rf{p1} as $\{
\left|B_{\alpha\beta}\right>\} = [B]'\{\left|\alpha\beta\right>\}$, where
\beq
	[B]' = \frac{1}{\sqrt{2}}\left[ \begin{array}{c c c c} 0 & 
	e^{i({\bf q}_g\delta{\bf r}_2 +\varphi_{g2} + kl_2)} & e^{i({\bf q}_g\delta{\bf r}_1 +\varphi_{g1} +kl_1)} & 0 \\
	e^{i({\bf q}_e\delta{\bf r}_2 +\varphi_{e2} + kl_2)} & 0 & 0 & e^{i({\bf q}_g\delta{\bf r}_1 +\varphi_{g1} + kl_1)}\\
	e^{i({\bf q}_e\delta{\bf r}_1 +\varphi_{e1} + kl_1)} & 0 & 0 & e^{i({\bf q}_g\delta{\bf r}_2 +\varphi_{g2} + kl_2)} \\
	0 & e^{i({\bf q}_e\delta{\bf r}_1 +\varphi_{e1} + kl_1)} & e^{i({\bf q}_e\delta{\bf r}_2 +\varphi_{e2} + kl_2)} & 0
\end{array}\right], \hspace{1cm}	\lb{00}
\eeq
 is a matrix of Bell operator $\hat{B}$. 

In general, the wave function $\left|B_{gg}\right>$ ($\left|B_{ge}\right>$) is not orthogonal to $\left|B_{ee}\right>$
($\left|B_{eg}\right>$). In order to make sure that all Bell's states are orthogonal one has to satisfy two conditions
\beqr
	0 & = & \left<B_{ee}|B_{gg}\right> = 
	W_{ge} \; e^{i(\varphi_{g1} - \varphi_{e2} + kl_1 - kl_2)} + 
	W_{eg}^* \; e^{i(\varphi_{g2} - \varphi_{e1} + kl_2 - kl_1)} \lb{7}\\
	0 & = & \left<B_{eg}|B_{ge}\right> = 
	W_{gg} \; e^{i(\varphi_{g1} - \varphi_{g2} + kl_1 - kl_2)} + 
	W_{ee}^* \; e^{i(\varphi_{e2} - \varphi_{e1} + kl_2 - kl_1)}, \lb{8}
\eeqr
where $W_{\alpha\beta} = e^{i({\bf q}_\alpha \delta{\bf r}_1 - {\bf q}_\beta \delta{\bf r}_2)}$. If the atoms are very
cold in a steep trap, so that they are deeply in the Lamb-Dicke regime, one should take $W_{\alpha\beta}=1$ in
Eqs.\rf{7} and \rf{8}. But we point out that the resulting conditions are actually independent of the atoms motion. Indeed,
\beq
	\left< W_{\alpha\beta}\right> = \left<W_{\alpha\beta}^* \right> = V_{\alpha}V_{\beta}
	\equiv 1 - D(T), \hspace{0.5cm}	 V_{\alpha} = \left<\sum_{n=0}^{-\infty}\frac{(-1)^n}{(2n)!}
	\left<({\bf q}_{\alpha}\delta{\bf r})^{2n}\right>_T\right>_{{\bf q}_{\alpha}}	\lb{10a}
\eeq
where $\left<...\right>_T$ and $\left<...\right>_{{\bf q}_{\alpha}}$ means, respectively, the average over the atom
 motion and over directions of registered photons. We average separately over symmetrical and statistically independent
 motion of each atom, drop index $i$ in $\delta {\bf r}_i$, and introduce parameter $D(T)$, $0 \leq D(T) < 1$, where $T$
 is the temperature associated with the random motion of the atoms.  One can see that $\left< W_{\alpha\beta} \right>$
 and $\left<W_{\alpha\beta}^* \right>$ disappear from orthogonality conditions \rf{7}, \rf{8}, which are reduced to a
 single condition
\beq
	\varphi_{g2}^0 - \varphi_{g1}^0 + \varphi_{e2}^0 - \varphi_{e1}^0 + 
	({\bf k}_e + {\bf k}_g)({\bf r}_2 - {\bf r}_1)  + 2 k(l_2-l_1) = \pi	\lb{10b}
\eeq
In our geometry we have $\varphi_{\alpha 1}^0 = \varphi_{\alpha 2}^0 $, and thus this condition becomes 
\beq
 	({\bf k}_e + {\bf k}_g)({\bf r}_2 - {\bf r}_1)  + 2 k (l_2-l_1) = \pi.	\lb{a1}
\eeq
There are various ways to fulfill this condition.  If one chooses ${\bf k}_e = - {\bf k}_g$, i.e., the $\sigma_+$ and
$\sigma_-$ lasers are propagating in opposite directions, the condition for orthogonality of the Bell's states is
obtained by adjusting the interferometer path difference so that $k (l_2-l_1) = \pi/2$. But it is also possible to take
${\bf k}_e = {\bf k}_g = {\bf k}_L$, together with ${\bf k}_L({\bf r}_2 - {\bf r}_1)= \pi/2$, obtained by adjusting the
trap's positions. Assuming then that $k l_2 = k l_1 = 2 n \pi$ for simplicity, taking (here and everywhere below) the
origin of the coordinate system on the atom 1 and defining ${\bf q} = {\bf k}_L - {\bf k}$, the Bell operator matrix can
be written:
\[
	[B]' = \frac{1}{\sqrt{2}}\left[ \begin{array}{c c c c} 
	0 & ie^{ i{\bf q}\delta{\bf r}_2} & e^{ i{\bf q}\delta{\bf r}_1} & 0 \\
	ie^{ i{\bf q}\delta{\bf r}_2} & 0 & 0 & e^{ i{\bf q}\delta{\bf r}_1} \\
	e^{ i{\bf q}\delta{\bf r}_1} & 0 & 0 & ie^{ i{\bf q}\delta{\bf r}_2} \\
	0 & e^{ i{\bf q}\delta{\bf r}_1} & ie^{ i{\bf q}\delta{\bf r}_2} & 0
\end{array}\right],
\]
which converts four initial atom states to four Bell states, which are orthogonal in average over the atom motion.
Though the condition for the orthogonality in average depends only on the atoms equilibrium positions, the final
fidelity of the conditional unitary transformation will obviously depend on the atoms motion, due to the $\delta{\bf
r}_1$ and $\delta{\bf r}_2$ in the $[B]'$ matrix.  

In order to simplify the local operations used in the rest of the paper, it is
convenient to perform two phase transformations for atom 2, that make the change $\left|e\right>_2
\rightarrow -i\left|e\right>_2$ just before the photon observation and $\left|e\right>_2 \rightarrow
i\left|e\right>_2$ right after it. Taking into account such transfornations as 
$diag\{1,-i,1-i\}[B]'diag\{1,i,1,i\} \equiv [B]$, where $diag$ means diagonal matrix, we find 
\beq
	[B] = \frac{1}{\sqrt{2}}\left[ \begin{array}{c c c c} 
	0 & -e^{ i{\bf q}\delta{\bf r}_2} & e^{ i{\bf q}\delta{\bf r}_1} & 0 \\
	e^{ i{\bf q}\delta{\bf r}_2} & 0 & 0 & e^{ i{\bf q}\delta{\bf r}_1} \\
	e^{ i{\bf q}\delta{\bf r}_1} & 0 & 0 & -e^{ i{\bf q}\delta{\bf r}_2} \\
	0 & e^{ i{\bf q}\delta{\bf r}_1} & e^{ i{\bf q}\delta{\bf r}_2} & 0
\end{array}\right], \hspace{1cm}	\lb{a2}
\eeq
which has real elements in the absence of atom motion $\delta{\bf r}_i = 0$. In a geometry where
the phases $\varphi_{\alpha i}^0$ can be independantly controlled, one can obtain the matrix
\rf{a2} more straightforwardly, for example by choosing in Eq.\rf{10b} $k l_2 = k l_1 = 2 n \pi$, ${\bf k}_e = {\bf
k}_g = {\bf k}_L$, ${\bf r}_1 = 0$, and 
\beq
	\varphi_{g1}^0 = \varphi_{e1}^0 = 0	\hspace{1cm}  \varphi_{e2}^0 = -{\bf k}_L{\bf r}_2,
	 \hspace{1cm} \varphi_{g2}^0 = \pi - {\bf k}_L{\bf r}_2.	\lb{b2}
\eeq
Below we refer to $[B]$ as a Bell operator matrix supposing either that condition \rf{a1} is true and the Bell operation
 is the photon observation procedure with the two phase transformations for atom 2, or that there is only the
 photon observation, but conditions \rf{b2} are satisfied.

In order to get a physical understanding about the quality of the Bell states preparation, we will now look in detail
whether the prepared states can violate Bell's inequalities.  In these calculations we will use the expression \rf{a2}
corresponding to ${\bf k}_e = {\bf k}_g = {\bf k}_L$, but similar results could be easily obtained by in the case where
${\bf k}_e = - {\bf k}_g$ (the fully phase-matched situation where the atoms' positions would cancel out is not
accessible with our experimental geometry).

\section{Bell's inequalities}\label{BI}

When the atoms are prepared in a Bell state, 
the statistical behavior of measurable quantities (such as
the population of state $\left|\alpha\right>_i$) is governed by the entangled wave-function
$\left|B_{\alpha\beta}\right>$. Here BI will be used as a simple experimental 
characterization of the degree of entanglement of the atom pair. 
As we will see below, either atoms motion or simultaneous excitation of the two atoms
may reduce or even suppress the BI violation.

In order to test BI we carry out the following sequence of operations

1. The atoms are prepared in one of states \rf{p1};

2. Atoms are excited by a weak laser pulses  under the conditions of Eq.\rf{a1}; 

3. One spontaneously emitted photon is registered. If there is no photon after some delay, the
    stages 1, 2 are repeated until one photon is registered;

4. Raman transitions for each atom are carried out so that
\beq
	\left|g\right>_i  \rightarrow \cos{(\theta_i)}\left|g\right>_i -  \sin{(\theta_i)}\left|e\right>_i
	\hspace{1cm} \left|e\right>_i \rightarrow \cos{(\theta_i)}\left|e\right>_i + 
	\sin{(\theta_i)}\left|g\right>_i; \lb{17}
\eeq

5. Populations of $\left|\alpha\right>_i$ states are measured;

6. Operations 1 -- 5 have to be repeated until a full statistical
    ensemble of results for the population of $\left|\alpha\right>_i$ states is obtained;

7. The steps 1 -- 6 are repeated for four different Raman transitions with four pairs of angles $\{\theta_1, \theta_2
    \}$, $\{\theta_1',\theta_2' \}$, $\{\theta_1', \theta_2 \}$, and $\{\theta_1,\theta_2'\}$.

After the operations 1 -- 6 are carried out, the state of atoms and a photon is $\left|{\bf
k}\right>\hat{R}\hat{B}\left|\alpha\beta\right>$, where the operator $\hat{R}$ describes Raman transitions \rf{17} for 
two atoms. The matrix $[RB]$ of $\hat{R}\hat{B}$ operator is the matrix product $[B][R(\theta_i)]$ where the matrix
$[R(\theta_i)]$ for the Raman transitions  is given by Eq.\rf{A2} of Appendix 2, and $[B]$ is given by
Eq.\rf{a2}. Here and below we denote the dependence on $\theta_1$ and $\theta_2$ as a dependence on $\theta_i$, when it
does not lead to confusions.

Let us call $P_{\alpha\beta}^{(\gamma\delta)}(\theta_i)$ the probability to find atoms in state $\left|\gamma\delta\right>$,
while the initial atom state is $\left|\alpha\beta\right>$. By taking the modulus square of each matrix element
in  $[RB]$ one can find
\[
	P_{gg}^{(gg)}(\theta_i)  = P_{ge}^{(ge)}(\theta_i) = P_{ge}^{(eg)}(\theta_i)  = P_{gg}^{(ee)}(\theta_i) = 
	0.5\left[ \sin^2{(\theta_1 - \theta_2)} + 0.5Q(T,\theta_i) \right]
\]\[
	 P_{ge}^{(gg)}(\theta_i)  = P_{gg}^{(ge)}(\theta_i) = P_{gg}^{(eg)}(\theta_i)  = P_{ge}^{(ee)}(\theta_i) = 
  	0.5\left[\cos^2{(\theta_1 - \theta_2)} -  0.5Q(T,\theta_i) \right], 
\]\beq
	P_{eg}^{(gg)}(\theta_i)  = P_{ee}^{(ge)}(\theta_i) = P_{ee}^{(eg)}(\theta_i)  = P_{eg}^{(ee)}(\theta_i) = 
 	0.5\left[\cos^2{(\theta_1 + \theta_2)} +   0.5Q(T,\theta_i) \right] \lb{19}
\eeq\[
	 P_{ee}^{(gg)}(\theta_i)  = P_{eg}^{(ge)}(\theta_i) = P_{eg}^{(eg)}(\theta_i)  = P_{ee}^{(ee)}(\theta_i) = 
  	0.5\left[ \sin^2{(\theta_1 + \theta_2)} -  0.5Q(T,\theta_i) \right],
\]
where $Q(T,\theta_i) = D(T)\sin{(2\theta_1)}\sin{(2\theta_2)}$ and $D(T)$ is given by Eq.\rf{10a}. One has $D(T) = 0$
in the absence of atoms motion,  in this case
the maximum violation of Bell's inequalities is obtained.  In the opposite (high temperature) situation, where
$D(T) \rightarrow 1$, one obtains from Eqs.\rf{19}
\beq
  P_{gg}^{(gg)}(\theta_i) \rightarrow 
 0.5\left[ \sin^2{(\theta_1)}\cos^2{(\theta_2)} +  \sin^2{(\theta_2)}\cos^2{(\theta_1)} \right],	\lb{19b}
\eeq
and similar expressions for the other probabilities. This corresponds to 
a ``classical" limit where BI cannot be violated. Thus $D(T)$ is a ``decoherence parameter" which grows up with the
temperature from $D(0)= 0$ to $\max{D} = 1$.

For each atom $i=1,2$ we define a random variable $\xi_i$, with values $+1$ or $-1$ depending on whether an atom is found,
respectively, in $\left|g\right>_i$ or $\left|e\right>_i$ state after registering a photon and carrying out the Raman
transition. With the help of Eqs.\rf{19} one can find $\left<\xi_1\right> = \left<\xi_2\right> = 0$,
$\left<\xi_1^2\right> = \left<\xi_2^2\right> = 1$, where the average is made over the results of a sequence of
operations 1 - 6. The correlation functions are given by: 
\[
	E_{\alpha\beta}(\theta_i) \equiv \frac{\left<\xi_1\xi_2\right> - \left<\xi_1\right>\left<\xi_2\right>}
	{ \sqrt{\left<\xi_1^2\right>\left<\xi_2^2\right>}} = \left<\xi_1\xi_2\right> =
	2[P_{\alpha\beta}^{(ee)}(\theta_i) - P_{\alpha\beta}^{(eg)}(\theta_i)], 
\]
so that 
\[
	E_{ge}(\theta_i) = - E_{gg}(\theta_i) = 
	\cos{ [2(\theta_1-\theta_2)] } - Q(T,\theta_i)   
\]\beq
	E_{eg}(\theta_i) = - E_{ee}(\theta_i) = 
	\cos{ [2(\theta_1+\theta_2)] } + Q(T,\theta_i).    \lb{20}
\eeq
As usual we define the quantity 
\beq
	S_{\alpha\beta}(\theta_i,\theta_i') = E_{\alpha\beta}(\theta_i) - E_{\alpha\beta}(\theta_1\theta_2') + 
	E_{\alpha\beta}(\theta_1'\theta_2) + 
	E_{\alpha\beta}(\theta_i')	\lb{21}
\eeq
for each initial state $ \left|\alpha,\beta\right>$, then the Bell's inequalities read \ct{Bell}
\beq
	-2 \leq S_{\alpha\beta}(\theta_i,\theta_i') \leq 2.	\lb{22}
\eeq
The results \rf{19} -- \rf{21} are similar to ones obtained for the BI test with polarization-entangled photon pairs
\ct{Thes}, the difference is that here the decoherence is taken into account by means of $D(T)$. In the next Section we
look for the violation of inequalities \rf{22} for each initial state of the two-atom system.
\section{Effect of atom motion on Bell's inequalities test}
In order to predict accurately the value of $S_{\alpha\beta}(\theta_i,\theta_i')$ we have to calculate the factor
$D(T)$, which depends on the trapping potential and the aperture angle $\theta_0$ of the input lens of the optical
system. In general, the trapping potential is an-harmonic, non-symmetric and $\theta_0$ is not small. All of these
complicates the precise calculation of $D(T)$, which will be carried out elsewhere.  Here we will consider
a simple order-of-magnitude estimation, using an harmonic approximation for the trapping potential. The
procedure carried out in Appendix 1 (see also \ct{harm_osc}) leads to
\beq
	D(T) \approx 1 - e^{ -T/T_{cr}},
	\hspace{1cm} k_B T_{cr} = \frac{h^2 \nu_{eff}^2}{2 E_R}, 		\lb{23}
\eeq
where $T_{cr}$ is a critical temperature such that $D(T > T_{cr}) \approx 1$, $\nu_{eff}$ is an effective frequency of
atom motion in the trap. For the aperture angle $\theta_0 = \pi/4$, as it is in our case, $\nu_{eff}^{-2} \approx
1.25\nu_{\perp}^{-2} + 0.75\nu_{||}^{-2}$, where $\nu_{\perp}$ and $\nu_{||}$ are the frequencies of the motion of atoms
in $x$, $y$ and in $z$ directions, respectively; $E_R = \hbar^2 k^2/(2 m_{at}) \equiv h \nu_R$ is the recoil energy, and
$k_B$ is Boltzmann constant.  Using $\nu_R = 3.6 kHz$ for $Rb^{87}$ atoms and our estimations $\nu_{\perp} = 200$~kHz
and $\nu_{||} = 50$~kHz we obtain $\nu_{eff} = 55$~kHz and $T_{cr} \approx 20~\mu$K. In general, $\nu_{eff}$ and
$T_{cr}$ depend on $\theta_0$ and the direction of the laser field. The maximum $T_{cr}$ is reached when ${\bf k}$ is
parallel to ${\bf k}_L$ for the most of the emitted photons.  For the geometrical arrangement displayed in Fig.2, the
variation of $T_{cr}$ as a function of $\theta_0$ is given by Eqs.\rf{B8} of Appendix 1, and it is displayed in Fig.3.

%
\begin{figure}[h]
\center{
\includegraphics[width=6cm]{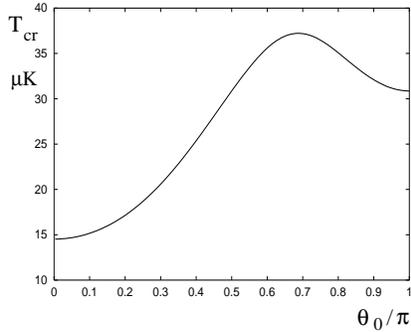} }\\
\caption{ \small Critical temperature as a function of the aperture angle of the optical system. }
\end{figure}
%
Let us choose parameters of Raman transitions 
\beq
	\theta_1 = 0, \hspace{0.5cm} \theta_2 = x,  \hspace{0.5cm} 
	\theta_1' = 2x, \hspace{0.5cm} \theta_2' = 3x,	\lb{24}
\eeq
and $T/T_{cr} = 0.5$, for such case factors $S_{ge}(x) = -S_{ee}(x)$ and $S_{eg}(x) = -S_{gg}(x)$ are shown in Fig.4a. 
%
\begin{figure}[h]
\center{
\includegraphics[width=5cm]{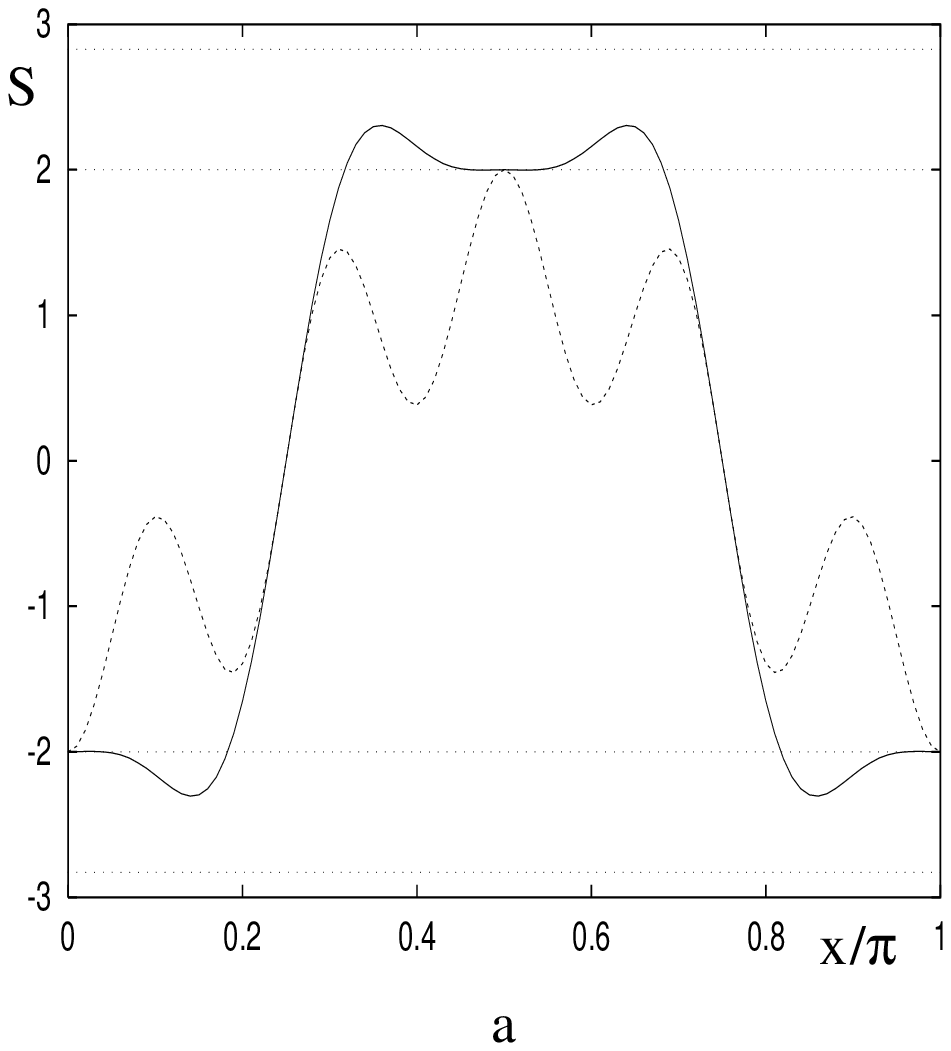} \hspace{1cm} \includegraphics[width=5cm]{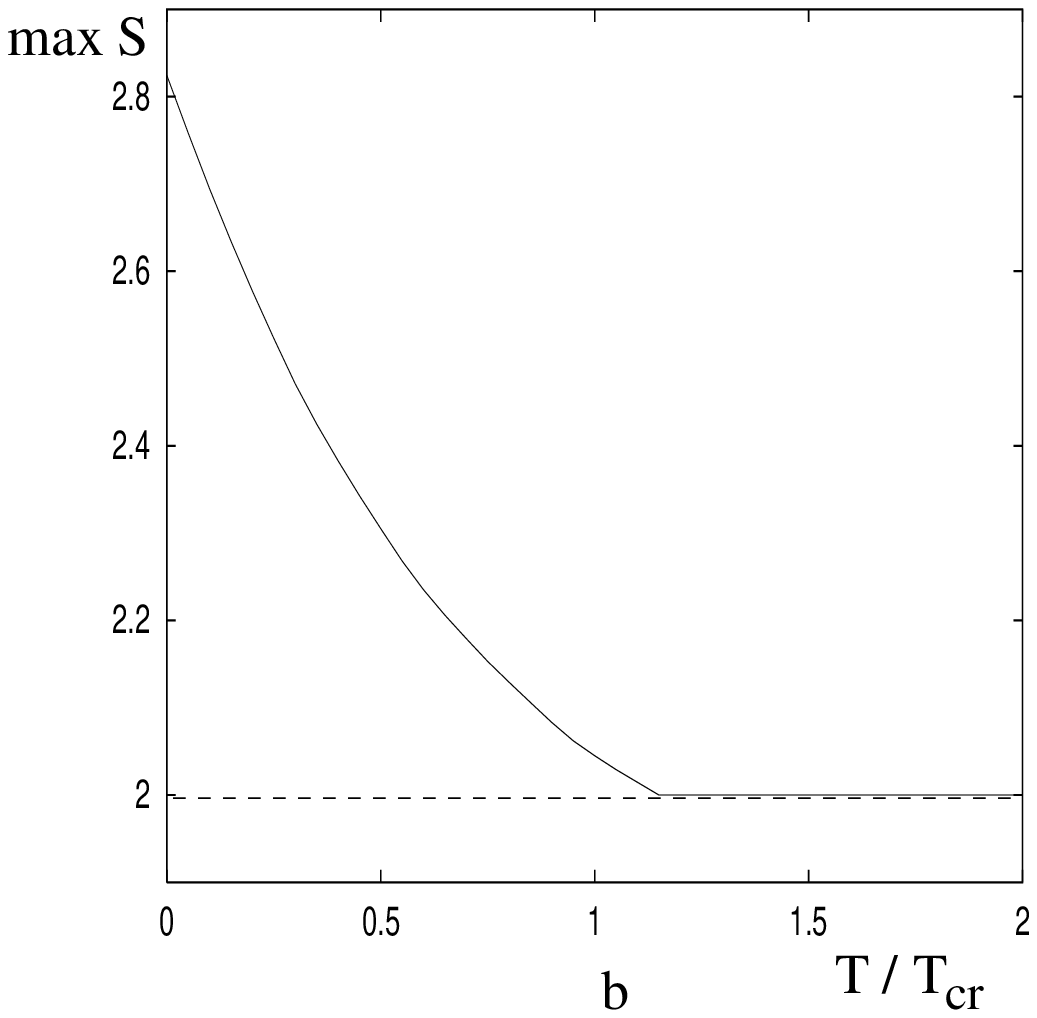}}\\
\caption{ \small (a): $S_{ge}(x) = -S_{ee}(x)$ (solid line) and $S_{eg}(x) = -S_{gg}(x)$ (dashed line) at $T/T_{cr} =
0.5$ and angles $\theta_1$, $\theta_2$, $\theta_1'$, $\theta_2'$ given by Eq.\rf{24}. Bell's inequalities are violated
 for $S_{ge}(x) = -S_{ee}(x)$. (b) The maxima of $|S_{\alpha\beta}|$ for different $T/T_{cr}$.  Dotted lines in
Fig.4a mark $|S| = 2$ and $|S| = 2\sqrt{2}$.}
\end{figure}
%
%
One can observe the violation of BI $|S_{\alpha\beta}(x)| > 2$ for $S_{ge}(x)$ and $S_{ee}(x)$, while
BI are satisfied for $S_{eg}(x)$ and $S_{gg}(x)$. This situation can be inverted by choosing $\theta_1
= 0$, $\theta_2 = -x$, $\theta_1' = 2x$, and $\theta_2' = -3x$, so that BI will be violated for
$S_{eg}(x)$ and $S_{gg}(x)$ but satisfied for $S_{ge}(x)$ and $S_{ee}(x)$. 
Therefore all four states do violate BI, but the combination of angles to be used depend on the state
in the pairwise fashion just described. 
Fig.4b shows the maxima of $|S_{\alpha\beta}|$ versus the normalized temperature of the atom motion found for 
 $\theta_{1,2}$, $\theta_{1,2}'$ given by Eqs.\rf{24}. The condition to violate BI for all four states 
(for suitable choices of Raman angles) is therefore that $T < T_{cr}$.
\section{Effect of multiple scattering on Bell's inequalities test.}
Now we take into account the excitation of two atoms together and examine how it influences the BI
violation. Let us first consider the case of only three levels in each atom shown in Fig.5.
%
%
\begin{figure}[h]
\center{
\includegraphics[width=13cm]{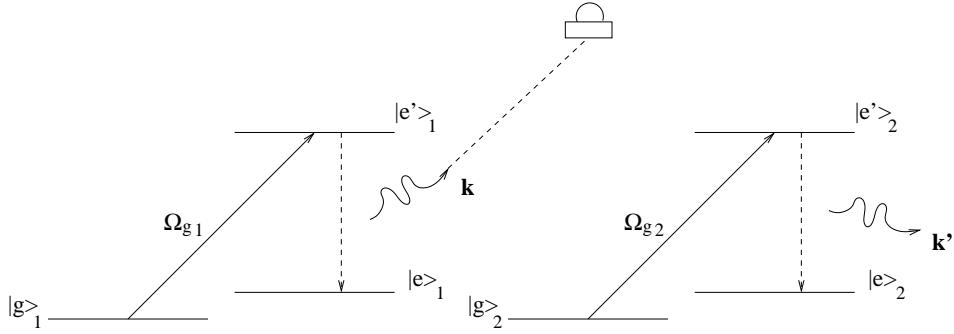} }\\
\caption{ \small The event when two atoms are excited together, both of them came to the $\left|e\right>$ state, but one
emitted photon is missed.}
\end{figure}
%
%
We examine a possibility that a photon emitted by one atom is registered, while another atom also emits a photon on
$\left|e'\right> \rightarrow \left|e\right>$ transition, but this photon is missed. After the excitation to
$\left|e'\right>_i$ state the wave function of atom $i = 1,2$ is given by Eqs.\rf{1}
\[
	\left|\Psi_g\right>_1 = a\left|g\right>_1 + b\left|e'\right>_1e^{i{\bf k}_L\delta{\bf r}_1}, \hspace{1cm}
	\left|\Psi_g\right>_2 = a\left|g\right>_2 - b\left|e'\right>_2e^{i{\bf k}_L\delta{\bf r}_2},	
\]
where we suppose that conditions \rf{b2} are satisfied.  If the photon $\bf k$ is emitted by one atom and registered,
  while another photon ${\bf k}'$ is emitted by the other atom and missed, then the atoms go from $\left|e'e'\right>$ to
  $\left|ee\right>$ state and we have:
\[
	e^{i{\bf k}_L(\delta{\bf r}_1 + \delta{\bf r}_2)}\left|e'e'\right> \rightarrow 
	e^{i{\bf k}_L(\delta{\bf r}_1 + \delta{\bf r}_2)}\left(e^{-i{\bf k}\delta{\bf r}_1}\left|ee'\right> + 
	e^{-i{\bf k}\delta{\bf r}_2}\left|e'e\right>\right)\left|{\bf k}\right>\rightarrow 
	f\left|ee\right>\left|{\bf k}{\bf k}'\right>, 
\]\beq
	f = e^{i({\bf q}\delta{\bf r}_1 + {\bf q'}\delta{\bf r}_2 + \varphi'_2)} + 
	e^{i({\bf q}\delta{\bf r}_2 + {\bf q'}\delta{\bf r}_1 + \varphi'_1)}, 	\lb{26}
\eeq
where ${\bf q}' = {\bf k}_L - {\bf k}'$, $\varphi'_{i}$ is the phase of a missed photon emitted by atom $i = 1,2$ and we
suppose, as usual, $kl_1 = kl_2 = 2\pi n$. It may also happen that
one atom emits the missed photon first, and then the registered
photon comes from another atom. In that case, one has to change the field state
$\left|{\bf k},{\bf k}'\right>$ in Eq.\rf{26} to $\left|{\bf k}',{\bf k}\right>$.  After registering a photon and
performing the phase transformation $\left|e\right>_2 \rightarrow i\left|e\right>_2$, the state $\left|gg\right>$ is projected to
the state
\beq
	\frac{1}{\sqrt{N}}\left[ \left(
	e^{i{\bf q}\delta{\bf r}_1} \left|eg\right> - e^{i{\bf q}\delta{\bf r}_2}\left|ge\right>\right)
	\left|{\bf k}\right> + (\xi)^{1/2} f\left|ee\right>
	\left(\left|{\bf k}',{\bf k}\right> + \left|{\bf k},{\bf k}'\right>\right) \right], \lb{28}
\eeq
where $\xi = (b/a)^2$.  Taking into account that the field state $\left|{\bf k}',{\bf k}\right>$ 
is orthogonal to $\left|{\bf k},{\bf k}'\right>$, and calculating
$\left<|f|^2\right>_T = 2$, one obtains the normalizing factor $N = 2[1 + 2\xi]$. 
 
After carrying out the Raman transitions, the atom states in the right part of Eq.\rf{28} are changed in accordance with
the transformation \rf{17}. Following the procedure of Section \ref{BI} we find
\beq
	E_{gg}(\theta_i) = - [1-D(T)] \sin{2\theta_1}\sin{2\theta_2}
	- \frac{\cos{2\theta_1}\cos{2\theta_2}}{1+2\xi}.	\lb{32}
\eeq
Fig.6 shows $S_{gg}(x)$ calculated with the help of Eqs.\rf{21},\rf{32} for $T/T_{cr} = 0.5$, $\theta_1$, $\theta_2$,
$\theta_1'$ and $\theta_2'$ given by Eqs.\rf{24} and various $\xi$. If the state $\left|e'\right>_i$ is excited by a
weak ``square" pulse, so that $\Omega_{gi}$ is constant during the excitation time and zero otherwise, then 
$\xi = |\Omega_{gi}|^2/\delta^2 \ll 1$, where $\delta$ is the detuning from the resonance on $\left|g\right>_i
\rightarrow \left|e'\right>_i$ transition. According with Fig.6, BI are still violated for
$|\Omega_{gi}|^2/\delta^2 \leq 0.15$.
%
%
\begin{figure}[h]
\center{
\includegraphics[width=6cm]{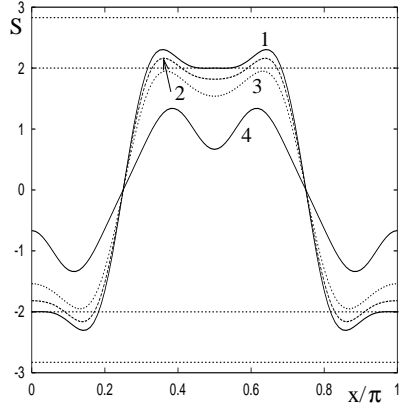} }\\
\caption{ \small Factor $S_{gg}(x)$ for Raman transition parameters given by Eq.\rf{24}, $T/T_{cr} = 0.5$ and
 $\xi = 0$ (curve 1), $\xi = 0.05$ (2), $\xi = 0.15$ (3) and $\xi = 1$ (4). BI violation is
possible for $\xi < 0.15$.}
\end{figure}
%
%

It is convenient to write the final two-atom state,
taking into account simultaneous excitation of two atoms for each initial state
$\left|\alpha\beta\right>$, under the following matrix form:
\beq
	\sqrt{2/N}\left(\left|{\bf k}\right>\hat{B} + 
	\left|{\bf k},\bf{k}'\right>\hat B^{(2)} \right)\left|\alpha\beta\right>,	\lb{32a}
\eeq
where the matrix $[B]$ of the operator $\hat{B}$ is given by Eq.\rf{a2}, the matrix of the operator $\hat B^{(2)}$ is 
\beq
 	[B^{(2)}] = \sqrt{2\xi }\left[ \begin{array}{c c c c} 0 & 0  & 0 & 1 \\
	0 & 0 & 1 & 0 \\ 0 & 1 & 0 & 0\\ 1 & 0 & 0 & 0	\end{array}\right], 	\lb{32b}
\eeq
and the state $\left|{\bf k},\bf{k}'\right>$ is orthogonal to $\left|{\bf k}\right>$ and normalized to 1.
\section{Bell's state measurement}
In the previous section we have shown that four orthogonal Bell's states can be prepared from four 
initial factorized states, under the condition of detecting a single photon. 
The reverse process, usually known as a Bell measurement,  is actually also realized using the same scheme. 
Here we will evaluate the efficiency of a whole sequence, including the preparation
followed by the measurement - two successive clicks will be therefore required.

By carrying out the steps 1 -- 3 of the procedure described in Section \ref{BI} we prepare a Bell state
$\left|B_{\alpha\beta}\right>$ of two atoms. Then $\left|B_{\alpha\beta}\right>$ can be projected to the pure state
$\left|\alpha\beta\right>$ of two atoms, or ``measured", by proceeding the steps 1 -- 3 with the phases of the
laser fields
\[
	\varphi_{e2}^0 = \pi - {\bf k}_L\Delta{\bf r}, \hspace{1cm} 
	\varphi_{g1}^0 = \varphi_{e1}^0 = 0, \hspace{1cm} \varphi_{g2}^0 =  - {\bf k}_L\Delta{\bf r}.
\]
Taking into account the multiple scattering, one
arrives to the final state of two atoms and spontaneously emitted photons after the Bell's state preparation from
$\left|\alpha\beta\right>$ state followed by the Bell's state measurement
\beq
	\sqrt{1/\tilde{N}}\left( \left|\tilde{\bf k}\right>\hat{\tilde{B}} + 
	\left|\tilde{\bf k},\tilde{\bf{k}}'\right>\hat{B}^{(2)} \right)
	\left(\left|{\bf k}\right>\hat{B} + 
	\left|{\bf k},\bf{k}'\right>\hat B^{(2)} \right)\left|\alpha\beta\right>,	\lb{M1}
\eeq
where $\tilde{N}$ is normalizing factor. A matrix $[\tilde{B}]$ of an operator $\hat{\tilde{B}}$ is $[B]^{-1}$ with
$\delta {\bf r}_i$ replaced by $\delta \tilde{\bf r}_i$, operator $\hat{B}^{(2)}$ is the same for the Bell's state
preparation and the measurement.  Because of the preparation and the measurement of a Bell's state are separated in
time, all field states in Eq.\rf{M1} are orthogonal to each other and the average over the atom motion $\left<\delta {\bf
r}_i\delta \tilde{\bf r}_j\right> = 0$, $i, j = 1,2$, since the atom motions on different time intervals are not
correlated.

The Bell's state measurement is not perfect due to the atom motion and the multiple scattering, so that the state
\rf{M1} is, in general, a linear combination of four states \rf{p1}. If a fidelity of the Bell's state measurement is
high, the probability to find atoms in $\left|\alpha\beta\right>$ initial state after the measurement approaches $1$,
while the probabilities to find any other atom states tends to $0$. A matrix for the transformation of the vector-column
of states \rf{p1} after the Bell's state preparation and the measurement is
\beq
	\sqrt{1/\tilde{N}} \left([B][\tilde{B}] + [B^{(2)}][\tilde{B}] + [B][{B}^{(2)}] + 
	[B^{(2)}]^2\right).	\lb{M2}
\eeq
Taking the square modulus of each element in the matrix \rf{M2} and calculating $\tilde{N} = [1 + 2\xi]^2$ one
converts matrix \rf{M2} to a matrix of probabilities to find atoms in $\left|\gamma\delta\right>$ final state starting
with $\left|\alpha\beta\right>$ initial state
\beq
 	\frac{1}{(1+2\xi)^2}\left[ \begin{array}{c c c c} 1 - f_1 + 4\xi^2 & 
	2\xi   & 2\xi  &  f_1  \\
	 2\xi  &  1 - f_1 + 4\xi^2  &  f_1  &  2\xi  \\ 
	 2\xi  &  f_1  & 1 - f_1 + 4\xi^2  &  2\xi 
	\\ f_1  &   2\xi  &  2\xi  & 1 - f_1 + 4\xi^2 	\end{array}\right],	\lb{M3}
\eeq\[
	 f_1(T) = D(T) - D^2(T)/2.
\]
In the case of perfect Bell's state preparation and the measurement the matrix \rf{M3} has diagonal elements equal to $1$
and other elements equal to $0$. Thus, we can take the diagonal element of the matrix \rf{M3} as the fidelity
$F_B(\xi,T)$ of Bell's state measurement
\beq
	F_B(\xi,T) =  1 - \frac{4\xi^2 + D(T) - D^2(T)/2}{(1 + 2\xi)^2}	\lb{M4}
\eeq
Fidelity $F_B$ is shown in Fig.7a as a function of $T/T_{cr}$ for
various $\xi$, it is shown in Fig.7b as a function of $\xi$ for various $T/T_{cr}$.
%
%
\begin{figure}[h]
\center{
\includegraphics[width=5cm]{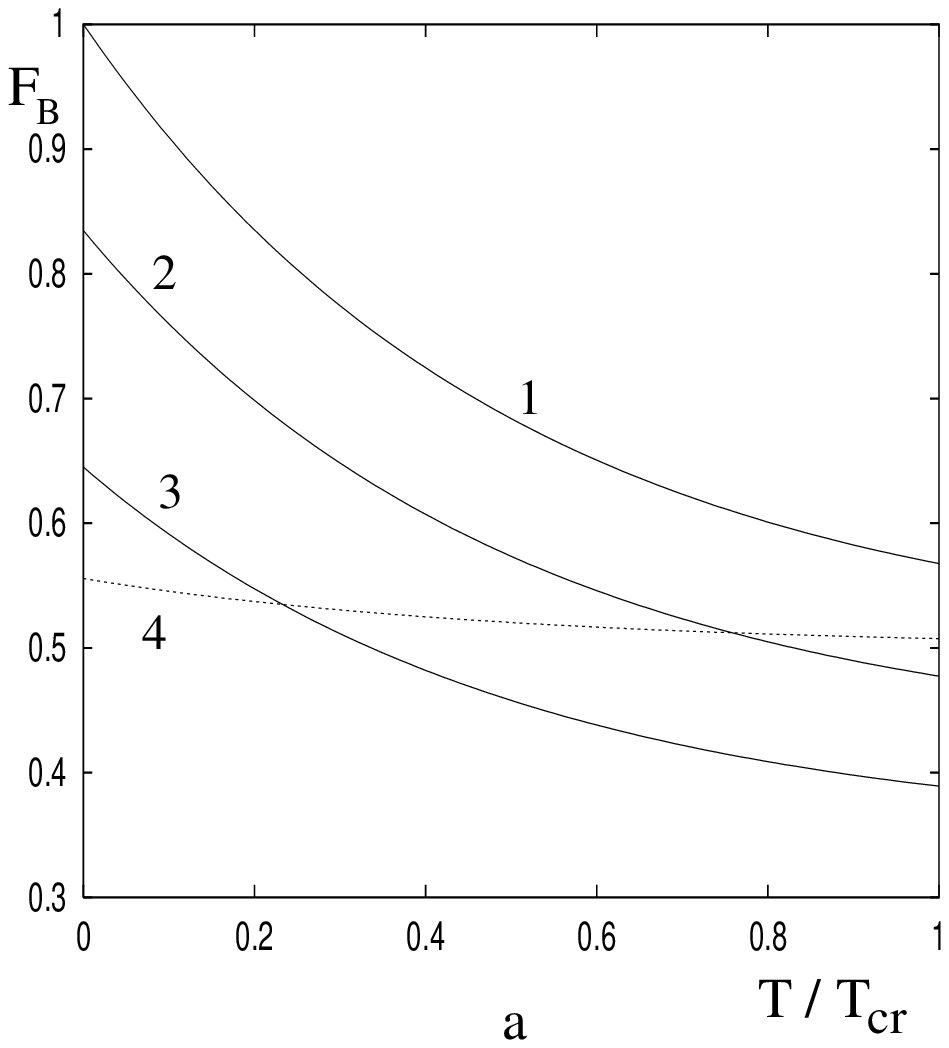} \hspace{1cm} \includegraphics[width=5cm]{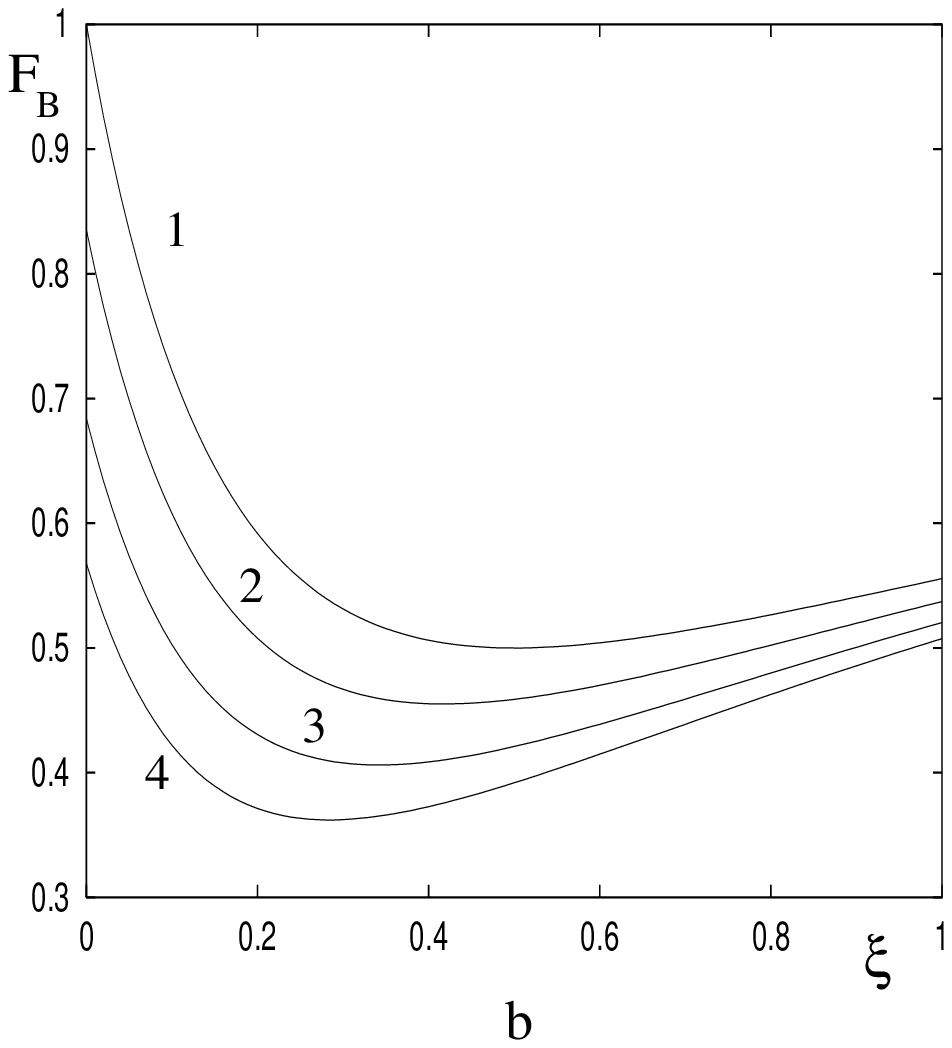}  }\\
\caption{ \small (a) Fidelity factor $F_B$ for Bell's state measurement as a function of $T/T_{cr}$ for $\xi = 0$ (curve
1), $\xi = 0.05$ (2), $\xi = 0.15$ (3) and $\xi = 1$ (4); (b) $F$ as a function of $\xi$ at $T/T_{cr} = 0$ (curve 1),
$T/T_{cr} = 0.2$ (2), $T/T_{cr} = 0.5$ (3), $T/T_{cr} = 1$ (4). }
\end{figure}
%
%
Small increase in $F_B$ for large $\xi \approx 1$ is because of the contribution of processes 
$\left|\alpha\alpha\right> \rightarrow \left|\beta\beta\right> \rightarrow \left|\alpha\alpha\right>$, $\alpha \neq
\beta$ grows up with $\xi$ due to the multiple photon scattering. However this is not so important for practical
cases, where $\xi \ll 1$.
\section{Quantum CNOT gate.}
We have shown so far that conditional Bell states preparation and measurement can be successfully achieved. 
Our result is actually more general than that, and shows that arbitrary ``conditional" unitary 
transformations on two qubits can be achieved by using  Raman rotations (applied locally to each
atom) and the detection of a single click. In order to demonstrate this we will
now show that the four Bell's states preparation  $\left|B_{\alpha\beta}\right> =
\hat{B}\left|\alpha\beta\right>$ can be turned into a``controlled-not" (CNOT) operation $\hat{C}$, described
by the matrix 
\beq
 	[C] = \left[ \begin{array}{c c c c} 1 & 0  & 0 & 0 \\
	0 & 1 & 0 & 0 \\ 0 & 0 & 0 & 1\\ 0 & 0 & 1 & 0	\end{array}\right]. 	\lb{32c}
\eeq
We prove that in our case $\hat{C} = \hat{H}_2\hat{B}\hat{H}_1$, where $\hat{H}_{1,2}$ are some local (single-atom)
operations and the matrix of Bell operation $\hat{B}$ is given by Eq.\rf{a2}. The CNOT operation can thus be realized
by the following procedure:

1. One of initial states \rf{p1} of atoms is prepared;

2. The local operation $\hat{H}_1$ is carried out;

3. Atoms are excited and a spontaneously emitted photon is registered. If there is no photons registered for a time $t_0
   \gg \Gamma^{-1}$ operations 1 - 3 has to be repeated;

4. The local operation $\hat{H}_2$ is carried out.

\noindent Let us suppose, for a while, that atoms do not move, so that the Bell operator is $\hat{B}_0$, which matrix
$[B_0]$ is given by Eq.\rf{a2} with $\delta {\bf r}_i = 0$. By definition $[C] = [H_1][B_0][H_2]$, where $[H_{1,2}]$ are
the matrices of local operations $\hat{H}_{1,2}$, and therefore 
\beq
	[H_1] = [C][H_2]^{-1}[B_0]^{-1}.	\lb{tr} 
\eeq
Taking the matrix $[H_2]^{-1}$ as a general local transformation for two-level atom, inserting it in
Eq.\rf{tr} with the requirement that 
the matrix product on the right of Eq.\rf{tr} should be a local transformation, we obtain
\beq
 	[H_1] = \frac{1}{2}\left[ \begin{array}{c c c c} i & i  & -i & -i \\
	-1 & 1 & 1 & -1 \\ i & i & i & i \\ -1 & 1 & -1 & 1	\end{array}\right], \hspace{1cm}
	[H_2] = \frac{1}{\sqrt{2}}\left[\begin{array}{c c c c} 0 & -1  & 0 & -i \\
	i & 0 & 1 & 0 \\ 0 & -1 & 0 & -i\\ -i & 0 & 1 & 0	\end{array}\right]. 	\lb{34}
\eeq
Details of the procedure of determining of $[H_{1,2}]$ are given in Appendix 2. As it can be seen from Eqs.  \rf{A9},
\rf{A10} of Appendix 2, operation $\hat{H}_1$ is the phase transformation $\left|g\right>_2 \rightarrow
i\left|g\right>_2$, after which the Raman transition \rf{17} with $\theta_1 = \pi/4$, $\theta_2 = -\pi/4$ is carried
out. Operation $\hat{H}_2$ starts with the Raman transition \rf{17} with $\theta_1 = -\pi/4$, $\theta_2 = -\pi/2$ after
which one makes the phase transformations $\left|e\right>_1 \rightarrow -i\left|e\right>_1$, $\left|g\right>_2
\rightarrow -i\left|g\right>_2$.

Now we take into account the atom motion, the simultaneous excitation of two atoms and find a fidelity of CNOT
operation. We suppose, that $\hat{H}_{1,2}$ transformations are much faster than a period of the atom motion in the trap,
in such case $\hat{H}_{1,2}$ does not depend at all on the atom motion. Indeed, by carrying out a fast local
operation with atom $i$, one can chose the origin of the coordinate system in that atom, which means ${\bf r}_{i0} +
\delta{\bf r}_0 \equiv 0$. Thus, using formula \rf{a2} with $\delta {\bf r}_i \neq 0$ and Eq.\rf{32a} we obtain an
operator $\hat{C}$ of a non-perfect CNOT transformation
\beq
	\hat{C} = \sqrt{2/N}\left(\left|{\bf k}\right>\hat{C}^{(1)} + 
	\left|{\bf k},{\bf k}'\right>\hat{C}^{(2)}\right),	\lb{35}
\eeq
where matrices of operators $\hat{C}^{(1,2)}$ are
\beq
	[C^{(1)}] = [H_1]\cdot[B]\cdot[H_2], \hspace{1cm} [C^{(2)}] = [H_1]\cdot[B^{(2)}]\cdot[H_2].	\lb{36}
\eeq
The matrices $[H_{1,2}]$ are given by Eqs.\rf{34}, and matrices $[B]$, $[B_2]$ are given by Eqs.\rf{a2}, \rf{32b},
respectively. 

We can build now matrices $[|C^{(1,2)}|^2]$, which elements are the square modulus of respective elements of
$[C^{(1,2)}]$, and calculate the matrix $[C_p]$ of probabilities to find atoms in $\left|\gamma\delta\right>$ 
state after CNOT operation, while $\left|\alpha\beta\right>$ was the initial atom state
\[
 	[C_p] = \frac{2}{N}\left\{[|C^{(1)}|^2] +[|C^{(1)}|^2]\right\} = 
	\frac{1}{2}\left[ \begin{array}{c c c c} 1+F & 1-F & 0 & 0 \\
	1-F & 1+F & 0 & 0 \\ 0 & 0 & 1-F & 1+F\\ 0 & 0 & 1+F & 1-F	\end{array}\right], 	
\]
where 
\beq
	F = \frac{1-D(T)}{1 + 2\xi}, 	\lb{37}
\eeq
$0 < F < 1$ is the fidelity of CNOT operation \rf{35}. Factor $F$ is shown in Fig.8a as a function of $T/T_{cr}$ for
various $\xi$, it is shown in Fig.8b as a function of $\xi$ for various $T/T_{cr}$.
%
%
\begin{figure}[h]
\center{
\includegraphics[width=5cm]{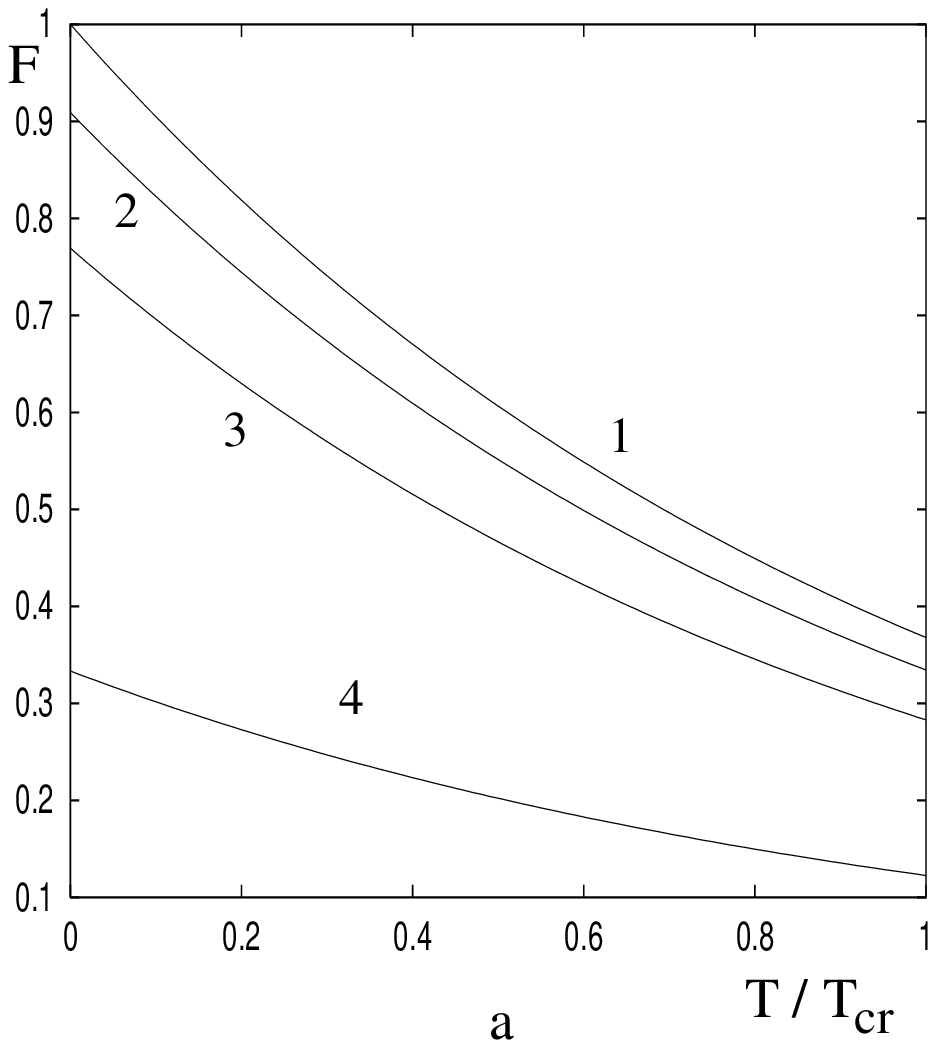} \hspace{1cm} \includegraphics[width=5cm]{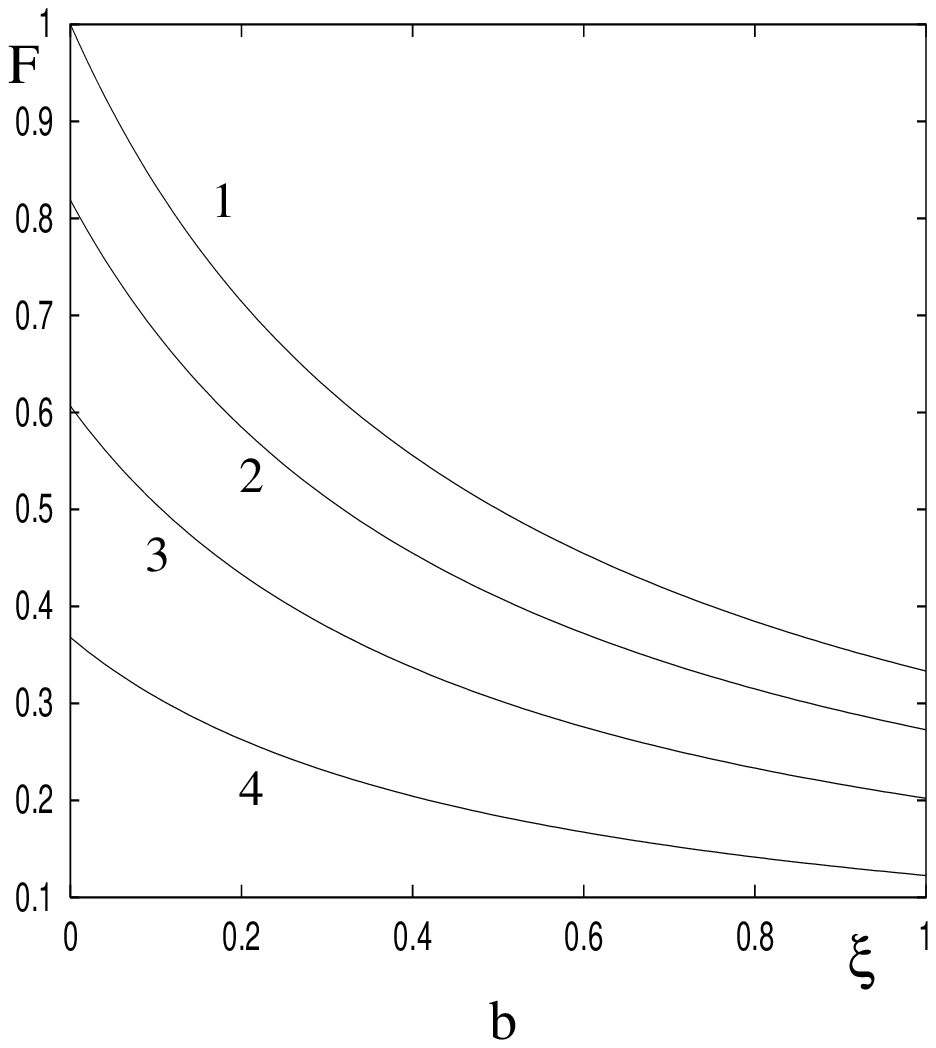}  }\\
\caption{ \small (a) Fidelity factor $F$ for quantum CNOT gate as a function of $T/T_{cr}$ for $\xi = 0$ (curve 1), $\xi
= 0.05$ (2), $\xi = 0.1$ (3) and $\xi = 1$ (4); (b) $F$ as a function of $\xi$ at $T/T_{cr} = 0$ (curve 1), $T/T_{cr} =
0.2$ (2), $T/T_{cr} = 0.5$ (3), $T/T_{cr} = 1$ (4). }
\end{figure}
%
%
\section{Discussion}

An important problem in the experimental demonstration of the conditional quantum CNOT gate operation is the suppression
of the atom motion, which can be done by cooling atoms in the traps up to temperatures of few $\mu$K, or by increasing
the atom oscillation frequencies. The last can be done, for example, by using standing-wave trapping fields, that
separates a trapped potential into several narrow wells with oscillation frequencies much higher than the
present 50 kHz obtained for the longitudinal motion in a tightly focussed beam.  Multiple
scattering gives rather small contribution, even for few percents of the atom excitation. 
An important characteristics of the fidelity
of the CNOT gate operation is the decoherence parameter D(T). For a reliable theoretical determination of D(T),
one needs to know accurately the trapping potential.
However, before doing experiments on BI violation, D(T) can also be determined experimentally by looking
at the interference fringes on the light emitted by the two atoms, irradiated on a closed transition \cite{harm_osc}.

Neutral atoms in a dipole trap are not the only candidates for implementing our conditional CNOT gate. In principle, the
gate can be realized with other resonant objects such as, for example, trapped ions,
or quantum dot molecules (QDM) incorporated in a solid
matrix \ct{Jap,QD_nature}. Each QDM consists in two closely positioned quantum dots with the ground state of each
dot split into two or more close states. The advantages of QDM are their fixed positions in
the matrix, and the possiblity to prepare the initial states electronically. The difficulty, however, is in providing the
coherence during the gate operation, which is quickly destroyed by the electron-phonon interaction.

The proposed simple scheme can be generalized straightforwardly to more complicated schemes with many elementary gates,
which may be called ``integrated conditional quantum logic blocks", or ICQLB. They can be constructed by the
increase of the number of atoms and (or) the number of ground states available in a single atom. There are $p^n$ initial
states of n atoms, if an identical photon can be emitted on transitions to $p$ different ground states.  However,
because of the photon observation process is not Hermitian, the maximum number $N(n,p)$ of obtained orthogonal Bell's
states is, in general, less than $p^n$, though $N(n,p)$ increases with $n$ and $p$. One can see, for example, that only
7 orthogonal Bell's states are possible for three atoms with the level scheme of Fig.1, for any choice of phases of laser
fields.  Determination of $N(n,p)$ is an important question for the theoretical modeling of ICQLB.

Another theoretical problem is how to find local transformations, which convert 
$N(n,p) \times N(n,p)$ matrix of the generalized 
Bell transformation, obtained by the photon observation procedure, to the matrix of
desirable logic transformation.  The procedure of Appendix 2 can be generalized, in principle, to higher-dimensional
cases, however it seems too cumbersome, and the development of a simpler procedure would be quite helpful.  We 
underline that even complicated ICQLB operates in the same five steps as the CNOT gate described above. The step 1
is the preparation of initial $N(n,p)$ states of atoms; 2 includes local transformations; 3 
is the excitation of atoms by weak resonant fields, repeated may be several times until a spontaneously emitted photon is
registered; 4 is another the local transformation; 5 is the determination of final
populations of atomic states. Each step can be carried out simultaneously for all atoms together, so that the operation
time of ICQLB is not much longer than for the elementary CNOT gate. The increase in the operation time of
complicated ICQLB can happen however, because the probability for $n$ atoms to emit more that one photon increases with $n$,
so that lower intensities of the exciting fields are required in order to avoid  multiple scattering.

As a conclusion, though we do not propose here a specific way to make the present scheme scalable,
it would be very interesting to study up to which point quantum computations
may be realized using conditional logical elements such as the ones described above.

\vskip 2cm
%
\section*{Appendix 1}
We consider an atom in the trapping potential as a harmonic oscillator, with a deviation from the equilibrium position given by
$\delta{\bf r}_i = \{\delta r_x,\delta r_y, \delta r_z\}$. Then $\left<e^{i{\bf q}\delta{\bf r}_i}\right>_T =
e^{-\left<({\bf q}\delta{\bf r}_i)^2\right>_T/2}$, which is the consequence that the thermal fluctuations of the
position of a harmonic oscillator are described by the Gaussian distribution function, and therefore
\beq
	 1- D(T) \equiv 
	\left<e^{i{\bf q}(\delta{\bf r}_1 - \delta{\bf r}_2)}\right> = \left<
	e^{-\left<({\bf q}\delta{\bf r})^2\right>_T}\right>_{\bf q}.	\lb{B2}
\eeq
In the coordinate system shown in Fig.2 $k_x =
k\sin{\theta}\cos{\varphi}$, $k_y = k\sin{\theta}\sin{\varphi}$, $k_z = k\cos{\theta}$ and $k_{Lx} \approx k$, $k_{Ly} =
k_{Lz} = 0$, so that
\beq
	\left<({\bf q}\delta{\bf r})^2\right>_T = 
	k^2\left[(1 - \sin{\theta}\cos{\varphi})^2\left<\delta r_x^2\right>_T  + 
	\sin^2{\theta}\sin^2{\varphi}\left<\delta r_y^2\right>_T  + 
	\cos^2{\theta}\left<\delta r_z^2\right>_T\right].		\lb{B3}
\eeq
For the one-dimensional quantum harmonic oscillator with the mass $m_{at}$, which oscillates along the
axes $\xi$ with the frequency $\omega_{\xi}$, $\xi = x,y,z$ in the thermal equilibrium
\beq
	\left<\delta r_{\xi}^2\right>_T  = 
	\frac{\hbar}{2m_{at}\omega_{\xi}}\coth{\left(\frac{\hbar\omega_{\xi}}{2k_BT}\right)} \approx 
	\frac{k_BT}{m_{at}\omega_{\xi}^2}, \lb{B4} 
\eeq
at $\hbar\omega_{\xi}/(k_BT) \ll 1$. Taking $\omega_x = \omega_y = 2\pi\nu_{\perp}$ and $\omega_z = 2\pi\nu_{||}$ we
obtain
\beq
	\left<({\bf q}\delta{\bf r})^2\right>_T = 
	k^2\frac{k_BT}{2\pi^2 m_{at}}\left[(1 + \sin^2{\theta} - 2\sin{\theta}\cos{\varphi})\frac{1}{\nu_{\perp}^2}  + 
	\cos^2{\theta}\frac{1}{\nu_{||}^2}\right].	\lb{B5}
\eeq	
The average 
\beq
	\left< u(\theta,\varphi) \right>_{\bf q} = 
	\int_{0}^{2\pi}d\varphi\int_{0}^{\theta_0}
	\sin{\theta}d\theta{\cal P}(\theta,\varphi) u(\theta,\varphi),	\lb{B5a}
\eeq
where  $u(\theta,\varphi)$ is some function and 
\beq
	{\cal P}(\theta,\varphi)  = C_0(1 - \sin^2{\theta}\cos^2{\varphi}) \lb{B6}
\eeq
is a probability that a photon, emitted to the solid angle of the optical system has a direction ${\bf k}$, $C_0$ is the
normalizing constant
\beq
	\frac{1}{C_0} = \int_{0}^{2\pi}d\varphi\int_{0}^{\theta_0}\sin{\theta}d\theta
	(1 - \sin^2{\theta}\cos^2{\varphi}) = 
	\frac{4\pi}{3}\left[1 - \frac{1}{4}(3\cos{\theta_0} + \cos^3{\theta_0})\right].	\lb{B7}
\eeq
Eq.\rf{B6} is obtained from the relation 
\[
	{\cal P}(\theta,\varphi) = C_0\sum_{\lambda = 1,2}[{\bf d}\cdot{\bf e}_{\lambda}({\bf k}/k)]^2 = 
	C_0[1 - ({\bf d}\cdot{\bf k})^2/k^2],
\]
where ${\bf e}_{\lambda}({\bf k})$ is the unit wave vector along one of two possible polarizations $\lambda = 1,2$
of a photon and ${\bf d}$ is a unit wave vector along the direction of the dipole momentum of atom transition directed
along axes $x$ as shown in Fig.2.

Thus, in order to find $D(T)$ one has to insert Eq.\rf{B5} into Eq.\rf{B2} and calculate $\left<...\right>_{\bf q}$ as
shown by Eq.\rf{B5a}. For a high input aperture of the optical system, that is our case, this procedure can hardly lead to 
an analytical result. For the estimations we use an approximation
\[
	\left<e^{-\left<({\bf q}\delta{\bf r})^2\right>_T}\right>_{\bf q} \approx
	e^{-\left<\left<({\bf q}\delta{\bf r})^2\right>_{T}\right>_{\bf q}},
\]
which is as better, as $D(T)$ is smaller. Thus, we arrive to $D(T) = 1 - e^{-T/T_{cr}(\theta_0)}$ where
\beq
	k_BT_{cr}(\theta_0) = \frac{h^2\nu_{eff}^2(\theta_0)}{2E_R}, \hspace{1cm}
	\frac{1}{\nu_{eff}^2(\theta_0)} = \frac{A_{||}(\theta_0)}{\nu_{||}^2} + 
	\frac{A_{\perp}(\theta_0)}{\nu_{\perp}^2}, \lb{B8} 
\eeq\[
	A_{\perp}(\theta_0) = 
	1 + \frac{4\pi C_0(\theta_0)}{5}\left( 1 - \frac{5\cos{\theta_0} - \cos^5{\theta_0}}{4}\right) \hspace{0.5cm}  
	A_{||}(\theta_0) = \frac{8\pi C_0(\theta_0)}{15}
	\left(1 - \frac{5\cos^3{\theta_0} + 3\cos^5{\theta_0}}{8}\right),	
\]
and $C_0(\theta_0)$ is determined by Eq.\rf{B7}.
\section*{Appendix 2}
General local transformations, which mixes the states $\left|g\right>_i$ and $\left|e\right>_i$ of the two-level atom $i =
1,2$, can be written in the form
\beq
	\left|g\right>_i  \rightarrow  (c_ie^{i\phi_{gi}}\left|g\right>_i -  
	s_ie^{i\phi_{ei}}\left|e\right>_i)e^{i\xi_{gi}} \hspace{1cm}
	\left|e\right>_i  \rightarrow  (c_ie^{i\phi_{ei}}\left|e\right>_i + 
	s_ie^{i\phi_{gi}}\left|g\right>_i)e^{i\xi_{ei}}.	\lb{A1} 
\eeq
They consist the phase transformation which changes the phase of the state $\left|\alpha\right>_i
\rightarrow e^{i\xi_{\alpha i}}\left|\alpha\right>_i$, the Raman transition given by Eq.\rf{17} with $c_i \equiv
\cos{(\theta_i)}$, $s_i \equiv \sin{(\theta_i)}$ and another phase transformation $\left|\alpha\right>_i
\rightarrow e^{i\phi_{\alpha i}}\left|\alpha\right>_i$. For our purposes, however, it is enough to consider
transformations \rf{A1} (or reverse to them) with $\phi_{\alpha i} = 0$.  The matrix of the local transformations carried
out with two atoms is, therefore
\beq 
	[L(\theta_i,\xi_{\alpha i})] = [M(\xi_{\alpha i})][R(\theta_i)] \equiv [M(\xi_{\alpha i})]
	\left[ \begin{array}{c c c c} 
	c_1c_2 & -c_1s_2 & -s_1c_2 & s_1s_2	\\ c_1s_2 & c_1c_2 & - s_1s_2 &  - s_1c_2\\
	s_1c_2 & -s_1s_2 & c_1c_2 & -c_1s_2 \\  s_1s_2 & s_1c_2 & c_1s_2 & c_1c_2 \end{array}\right],\lb{A2}
\eeq
where
\beq
	[M(\xi_{\alpha i})] \equiv [M(\xi_{g1}, \xi_{e1}, \xi_{g2}, \xi_{e2})]  = 
	{\rm diag} \{ e^{i\xi_{gg}}, e^{i\xi_{ge}}, e^{i\xi_{eg}}, e^{i\xi_{ee}}\}, \hspace{1cm}
	\xi_{\alpha\beta} \equiv \xi_{\alpha 1} + \xi_{\beta 2},	\lb{AA2} 
\eeq
is the diagonal matrix of the phase transformation and $[R(\theta_i)] \equiv [R(\theta_1,\theta_2)]$ is the matrix of
the Raman transformation \rf{17}.

 Let us take $[H_2]^{-1} = [L(\theta_i,\xi_{\alpha i})]$, insert it into
 Eq.\rf{tr} and find
\beq
	[H_1] = [M_1(\xi_{\alpha i})]\left[ \begin{array}{c c c c} 
	c_1s_2 - s_1c_2 & c_1c_2 + s_1s_2 & c_1c_2 - s_1s_2 & 
	- c_1s_2 - s_1c_2\\ -c_1c_2 - s_1s_2  & c_1s_2 - s_1c_2 & c_1s_2 + s_1c_2 &  
	c_1c_2 - s_1s_2 \\ c_1s_2 - s_1c_2 & s_1s_2 + c_1c_2 & s_1s_2 - c_1c_2 & 
	s_1c_2 + c_1s_2 \\  s_1s_2 + c_1c_2 & s_1c_2 - c_1s_2 & s_1c_2 + c_1s_2 & 
	c_1c_2 -  s_1s_2 \end{array}\right], \lb{A3}
\eeq
where $[M_1(\xi_{\alpha i})]$ is a diagonal matrix obtained by interchanging two last elements of 
$[M(\xi_{\alpha i})]$.

Our goal is to determine $\theta_i$ and $\xi_{\alpha i}$, such that the matrix given by Eq.\rf{A3} can be
represented as
\beq
	[H_1] =	[L(\tilde{\theta}_i,\tilde{\xi}_{\alpha i})] \lb{A4}
\eeq
with some $\tilde{\theta}_{1,2}$, $\tilde{\xi}_{\alpha i}$.  By comparing the matrices given by Eq.\rf{A2} and \rf{A3} we
see, that Eq.\rf{A4} can be true only if $|c_i| = |s_i|$, that is when $\theta_i = \pm \pi/4$, $\theta_i = \pm
3\pi/4$; or when $c_i = 0$ or $s_i = 0$, while $|c_j| = |s_j|$, $j \neq i$. We chose $c_2 = 0$, $s_2 = 1$, that is for
$\theta_2 = \pi/2$ and $c_1 = s_1$ that is for $\theta_1 = \pi/4$, so that
\beq
	[H_1] = \left[ \begin{array}{c c c c} 
	e^{i\xi_{gg}} & e^{i\xi_{gg}} & - e^{i\xi_{gg}} & -e^{i\xi_{gg}}\\ 
	- e^{i\xi_{ge}} & e^{i\xi_{ge}} & e^{i\xi_{ge}} &  - e^{i\xi_{ge}}\\
	e^{i\xi_{ee}} & e^{i\xi_{ee}} & e^{i\xi_{ee}} & e^{i\xi_{ee}} \\  
	e^{i\xi_{eg}} &  -e^{i\xi_{eg}} & e^{i\xi_{eg}} & -e^{i\xi_{eg}} 
	\end{array}\right]. \lb{A7}
\eeq
We can see now, that the matrix $[L(\tilde{\theta}_i,\tilde{\xi}_{\alpha i})]$ is very similar to the matrix
\rf{A7} if we take $\tilde{\theta}_1 = \pi/4$ and $\tilde{\theta}_2 = -\pi/4$, so that 
\beq
	[L(\pi/4,-\pi/4, \tilde{\xi}_{\alpha i})] = \left[ \begin{array}{c c c c} 
	e^{i\tilde{\xi}_{gg}} & e^{i\tilde{\xi}_{gg}} & - e^{i\tilde{\xi}_{gg}} & -e^{i\tilde{\xi}_{gg}}\\ 
	- e^{i\tilde{\xi}_{ge}} & e^{i\tilde{\xi}_{ge}} & e^{i\tilde{\xi}_{ge}} &  - e^{i\tilde{\xi}_{ge}}\\
	e^{i\tilde{\xi}_{eg}} & e^{i\tilde{\xi}_{eg}} & e^{i\tilde{\xi}_{eg}} & e^{i\tilde{\xi}_{eg}} \\  
	-e^{i\tilde{\xi}_{ee}} &  e^{i\tilde{\xi}_{ee}} & -e^{i\tilde{\xi}_{ee}} & e^{i\tilde{\xi}_{ee}} 
	\end{array}\right]	 \lb{A8}
\eeq
with 
\beq
	 \tilde{\xi}_{\alpha\beta} \equiv  \tilde{\xi}_{\alpha 1} + \tilde{\xi}_{\beta 2}.	\lb{A8a}
\eeq
Apart of the notations for phases, the only difference between the matrices in Eqs.\rf{A8} and Eqs.\rf{A7} is the
opposite signs of the elements in the last raws. The simplest way to eliminate this difference by choosing
$\tilde{\xi}_{ee} = \pi$ is not permitted by relations \rf{A8a}. However, by the examination of Eqs.\rf{A7} and \rf{A8}
one can see that they are equivalent for $\xi_{eg} = \pi$, $\xi_{ee} = \xi_{gg} = \pi/2$, $\xi_{ge} = 0$ and
$\tilde{\xi}_{eg} = \tilde{\xi}_{gg} = \pi/2$, $\tilde{\xi}_{ge} = \tilde{\xi}_{ee} = 0$. Such choice does not
contradict with Eqs.\rf{AA2}, \rf{A8a}, it corresponds to $\xi_{g1} = 0$, $\xi_{g2} = \pi/2$, $\xi_{e1} = \pi/2$ ,
$\xi_{e2} = 0$ and $\tilde{\xi}_{g1} = 0$ $\tilde{\xi}_{g2} = \pi/2$, $\tilde{\xi}_{e1} = \tilde{\xi}_{e2} =
0$. Inserting such values of $\tilde{\xi}_{\alpha\beta}$ and $\tilde{\theta}_1 = \pi/4$, $\tilde{\theta}_2 = -\pi/4$
into Eqs.\rf{A8}, \rf{A4} we find
\beq
	[H_1] = [M(0,0,\pi/2,0)][R(\pi/4,-\pi/4)]. \lb{A9}
\eeq
Otherwise, the matrix $[H_2]^{-1}$ is, by definition, given by Eqs.\rf{A2}. Inserting there $\theta_1
= \pi/4$, $\theta_2 = \pi/2$ and the values of $\xi_{\alpha i}$ given above we arrive to 
\beq
	[H_2] =[R(-\pi/4,-\pi/2)][M(0,-\pi/2,-\pi/2,0)]. \lb{A10}
\eeq
where we take into account that $[R(\pi/4,\pi/2)]^{-1} = [R(-\pi/4,-\pi/2)]$ and $[M(0,\pi/2,\pi/2,0)]^{-1} =
[M(0,-\pi/2,-\pi/2,0)]$. Matrices $[H_{1,2}]$ are given explicitly by Eqs.\rf{34}. Obviously, that matrices given by
Eqs.\rf{A9}, \rf{A10} are not the only ones which satisfy Eq.\rf{tr}. However other possible local transformations will
be similar to the ones given by matrices \rf{A9}, \rf{A10}.\\

{\bf ACKNOWLEDGEMENTS} This work was supported by the European IST/FET project `QUBITS' and by the European IHP network
`QUEST'. I.~Protsenko is also grateful to Russian Foundation for Basic Research, grant 01-02-17330, for support.\\

\end{document}